\documentstyle[referee]{laa}

\begin{document}
\thesaurus{07  
	   (07.09.1; 
	   07.13.1;  
	   07.13.2   
	   )}

\title{Meteor Streams of Comet Encke. \\
Taurid Meteor Complex.}

\author{ Jozef Kla\v{c}ka }
\institute{Institute of Astronomy,
Faculty for Mathematics and Physics,
Comenius University, \\
Mlynsk\'{a} dolina,
842 15 Bratislava,
Slovak Republic}
\date{}
\maketitle
\begin{abstract}
Application of the theoretical results of the author are presented for the case
of meteor streams of comet Encke. Theoretical determination of meteor orbit for
comet Encke, as a parent body, is presented.
Four significant theoretical meteor streams, corresponding to Tauds N, Tauds S,
$\beta$ Tauds and $\xi$ Perds are found.
The meteor stream membership criterion is applied to the photographic
orbits of The IAU Meteor Data Center in Lund:
Taurid meteor stream is found for several possible areas in phase-space
of orbital elements.

\end{abstract}

\section{Introduction}
Application of the meteor stream membership criterion and the method
of prediction of theoretical radiants, suggested in
Kla\v{c}ka (1995a: Meteor stream membership criteria - paper I) and
Kla\v{c}ka (1995b: Meteor streams and parent bodies -- paper II),
is presented for the case
of meteor streams connected with the comet Encke.

Gravitational forces acting on meteoroids of the comet Encke
yield that semimajor axis $a$ (energy $E$, $E = 1 / a$ -- the sign is not
important, now),
$z-$component of the angular momentum
( $L_{z}$ $\equiv H_{z}$ $=$ $\sqrt{a (1 - e^{2})}$ $ \cos i$ )
and the angle $\pi$ ($\pi = \omega ~+~ \Omega$)
show small dispersions during evolution for several 10,000 years.
This suggest that these quantities should be used in determining the meteor
stream membership and in prediction of theoretical radiants.
$e$ -- eccentricity was taken as the fourth important quantity
for determination of meteor stream.
The reason for this choice is that distributions in $E$, $L_{z}$,
$\pi$ and $e$ may be approximated with normal distribution. The data
for the photographic meteor orbits of The
IAU Meteor Data Center in Lund were used.

\section{Theoretical prediction of radiants}
Let us consider that comet Encke is the only parent body for a meteor
stream. We use the result of section 6 of paper I in the form of
``Energy $- H_{z} - \pi$ method'' (see section 5.1 in paper II) for the
purpose of finding of meteor stream(s). The local
minima yield four theoretical radiants, summarized in Table 1; the theoretical
radiants should correspond to $\beta$ Tauds, Tauds S, Tauds N and
$\xi$ Perds.

\begin{table}
\caption{Theoretical radiants for comet Encke as a parent body.
Energy-H$_{z}$-$\pi$ method.}
\begin{flushleft}
\begin{tabular}{rrrrrrrrr}
\hline
\hline
$q [AU]$ & $e$ & $i [^{\circ}]$ & $\omega [^{\circ}]$ & $\Omega [^{\circ}]$ & $\alpha [^{\circ}]$
& $\delta [^{\circ}]$ & $\lambda_{\odot} [^{\circ}]$ & $| \vec{\bigtriangleup H} |^{2}$ \\
\hline
0.326 & 0.853 & 6.6 & 242  &  278 &  84 & 18 &	98 & 0.009 \\
0.324 & 0.854 & 4.8 & 118  &   42 &  54 & 16 & 222 & 0.011 \\
0.321 & 0.855 & 0   & 298  &  222 &  52 & 19 & 222 & 0.014 \\
0.321 & 0.855 & 0   &  62  &   98 &  85 & 24 &	98 & 0.015 \\
\hline
\hline
\end{tabular}
\end{flushleft}
\end{table}

\section{Photographic data and membership to meteor stream}
Section 5 of paper I is used in determining the membership to meteor stream.
The data for the photographic meteor orbits of The
IAU Meteor Data Center in Lund were used. We have found that orbital
elements $E$, $L_{z}$, $\pi$ and $e$ may be approximated with normal
distribution. Thus, the density function in Eq. (8) of paper I is taken
in the form of four-dimensional normal density function. The area $\Omega$
in Eq. (8) of paper I may be chosen in an infinity ways -- we take, in this
paper, only some of special cases summarized by the form
\begin{equation}\label{1}
D^{p} \equiv \sum_{i=1}^{4}
\left ( \frac{| Q_{i} ~-~ Q_{iM} |}{r_{i}~ \sigma_{i}} \right )^{p} \le
D_{c} ^{p} ~,
\end{equation}
where $Q_{1} = E$, $Q_{2} = e$, $Q_{3} = H_{z}$, $Q_{4} = \pi$, and,
the value $D_{c}$ corresponds to the case when multidimensional normal
distribution yields that meteors with $D > D_{c}$ are members of a group,
for given $r_{i}$ ($i$ = 1 to 4) and $p$, only with probability less than 5 $\%$
($Q_{iM}$ corresponds to the mean value of the quantity $Q_{i}$).
We consider the following cases, as examples: \\
1. $p =$ 2, $r_{1} = r_{2} = r_{3} = r_{4} =$ 1. \\
2. $p =$ 1, $r_{1} = r_{2} = r_{3} =$ 1, $r_{4} \rightarrow \infty$, \\
3. $p =$ 2, $r_{1} = r_{2} = r_{3} =$ 1, $r_{4} \rightarrow \infty$, \\
(We must stress that Eq. (1) is principally different from Eq. (1) in
Kla\v{c}ka (1992) -- the case 1 corresponds, roughly, to Eq. (1) in
Kla\v{c}ka (1992); however, even in the case that area $\Omega$ is of the
type of a quadratic form, it may be even of indefinite type.)

\subsection{Description of the procedure and basic results}
As we have already mentioned, the data for the photographic meteor orbits
of The IAU Meteor Data Center in Lund were used. The requirement
0.15 AU $\le q \le$ 0.5 AU and 0.4 $\le e \le$ 0.999 and $i \le 20^{\circ}$
yields 418 objects of the total set. The set of 418 meteors was used in
finding the required meteor stream.

The case 1, already defined, was ellaborated in the following way.
At first, 418/36 $=$ 11.6
objects per 10$^{\circ}$. The interval of $\pi$, in which
number of objects per 10$^{\circ}$ is greater than 11.6, corresponds to
$\pi \in ( 110^{\circ}, 190^{\circ}) \equiv I$ (no object exists for
$\pi \in ( 105^{\circ}, 110^{\circ})$). The distribution for $\pi \in I$
may be approximated by normal. Thus, initial set of 418 objects reduced
to 213 objects with $\pi \in I$.
Four-dimensional normal distribution
was used (quantities
$Q_{1} = E$, $Q_{2} = e$, $Q_{3} = H_{z}$, $Q_{4} = \pi$). Since only several
objects had $8.5^{\circ} < i < 20^{\circ}$ (no object existed for
$i \in$ ( 8.5$^{\circ}, 9.5^{\circ}$)),
we reduced the initial set
of 213 objects into a set with $i < 8.5^{\circ}$ -- the
final starting set contained 191 objects.
Four-dimensional normal distribution
was applied to the set of 191 objects (quantities
$Q_{1} = E$, $Q_{2} = e$, $Q_{3} = H_{z}$, $Q_{4} = \pi$). The final result
is characterized by 4-dimensional normal distribution with mean values
$Q_{1M}$, $Q_{2M}$, $Q_{3M}$, $Q_{4M}$ and covariant matrix of the vector
($Q_{1}$, $Q_{2}$, $Q_{3}$, $Q_{4}$) with components $\sigma_{ij}$
($i$, $j$ $=$ 1 to 4) -- see Table 2.  We have
obtained 130 objects for the case 1
(probability 95 $\%$ -- $D_{c} =$ 3.31).
(Quantities $\sigma_{i}$, $i =$ 1 to 4, used in Eq. (1) are given
by the relations $\sigma_{i} = \sqrt{\sigma_{ii}}$, $i =$ 1 to 4.)

\begin{table}
\caption{Mean values and components of covariant matrix for the case 1.
(Probability 95 $\%$ -- $D_{c} =$ 3.31).}
\begin{flushleft}
\begin{tabular}{cccc}
\hline
\hline
   $Q_{1M}$   &    $Q_{2M}$   &    $Q_{3M}$   &    $Q_{4M}$   \\
\hline
  0.4701585   &   0.8334692   &   0.8028469   & 154.4876923   \\
\hline
\hline
$\sigma_{11}$ & $\sigma_{12}$ & $\sigma_{13}$ & $\sigma_{14}$  \\
\hline
   0.0036200  &   -0.0006914  &   -0.0015446  &   -0.4048438   \\
\hline
\hline
$\sigma_{21}$ & $\sigma_{22}$ & $\sigma_{23}$ & $\sigma_{24}$  \\
\hline
  -0.0006914  &    0.0007624  &   -0.0010880  &   -0.0410647   \\
\hline
\hline
$\sigma_{31}$ & $\sigma_{32}$ & $\sigma_{33}$ & $\sigma_{34}$  \\
\hline
  -0.0015446  &   -0.0010880  &    0.0037522  &    0.4382409   \\
\hline
\hline
$\sigma_{41}$ & $\sigma_{42}$ & $\sigma_{43}$ & $\sigma_{44}$  \\
\hline
  -0.4048438  &   -0.0410647  &    0.4382409  &  278.4472892   \\
\hline
\hline
\end{tabular}
\end{flushleft}
\end{table}

In the cases 2 and 3, already defined, the three-dimensional normal distribution
was used (quantities
$Q_{1} = E$, $Q_{2} = e$, $Q_{3} = H_{z}$).
In both cases the statistics yielded that the obtained
distributions in $\pi$ corresponded to normal for $\pi \in ( 110^{\circ},
190^{\circ}) \equiv I$, the number of objects with $\pi \in I'$ was smaller
than the number of objects with $\pi \in I$. As for the distribution
in inclinations, the same holded as in the case 1. Thus, the final starting
set of the data contained 191 objects.
Three-dimensional normal distributions
were applied to the set of 191 objects (quantities
$Q_{1} = E$, $Q_{2} = e$, $Q_{3} = H_{z}$). The final results
are characterized by 3-dimensional normal distributions with mean values
$Q_{1M}$, $Q_{2M}$, $Q_{3M}$ and covariant matrix of the vector
($Q_{1}$, $Q_{2}$, $Q_{3}$) with components $\sigma_{ij}$
($i$, $j$ $=$ 1 to 3) -- see Table 3 for the case 2 and Table 4 for the case 3.
We have
obtained 130 objects for the case 2
(probability 95 $\%$ -- $D_{c} =$ 4.69) and 131 objects for the case 3
(probability 95 $\%$ -- $D_{c} =$ 3.02).
(Quantities $\sigma_{i}$, $i =$ 1 to 3, used in Eq. (1) are given
by the relations $\sigma_{i} = \sqrt{\sigma_{ii}}$, $i =$ 1 to 3.)

\begin{table}
\caption{Mean values and components of covariant matrix for the case 2.
(Probability 95 $\%$ -- $D_{c} =$ 4.69).}
\begin{flushleft}
\begin{tabular}{ccc}
\hline
\hline
   $Q_{1M}$   &    $Q_{2M}$   &    $Q_{3M}$	\\
\hline
  0.4672623   &   0.8323077   &   0.8076754	\\
\hline
\hline
$\sigma_{11}$ & $\sigma_{12}$ & $\sigma_{13}$	\\
\hline
   0.0034363  &   -0.0006934  &   -0.0014475	\\
\hline
\hline
$\sigma_{21}$ & $\sigma_{22}$ & $\sigma_{23}$	\\
\hline
  -0.0006934  &    0.0007924  &   -0.0011204	\\
\hline
\hline
$\sigma_{31}$ & $\sigma_{32}$ & $\sigma_{33}$	\\
\hline
   -0.0014475 &   -0.0011204  &    0.0037191	\\
\hline
\hline
\end{tabular}
\end{flushleft}
\end{table}

\begin{table}
\caption{Mean values and components of covariant matrix for the case 3.
(Probability 95 $\%$ -- $D_{c} =$ 3.02).}
\begin{flushleft}
\begin{tabular}{ccc}
\hline
\hline
   $Q_{1M}$   &    $Q_{2M}$   &    $Q_{3M}$	\\
\hline
  0.4676122   &   0.8334656   &   0.8049176	\\
\hline
\hline
$\sigma_{11}$ & $\sigma_{12}$ & $\sigma_{13}$	\\
\hline
   0.0035460  &   -0.0006960  &   -0.0015081	\\
\hline
\hline
$\sigma_{21}$ & $\sigma_{22}$ & $\sigma_{23}$	\\
\hline
  -0.0006960  &    0.0007820  &   -0.0011135	\\
\hline
\hline
$\sigma_{31}$ & $\sigma_{32}$ & $\sigma_{33}$	\\
\hline
   -0.0015081 &   -0.0011135  &    0.0037759	\\
\hline
\hline
\end{tabular}
\end{flushleft}
\end{table}

\subsection{Comparison of the results}
Table 5 presents objects classified as members of all three meteor streams
-- each object belongs to the stream in the cases 1,
2 and 3, simultaneously (127 objects).

\begin{table}
\caption{Objects belonging to all three meteor streams (all three cases).}
\begin{flushleft}
\begin{tabular}{cccccccccc}
\hline
\hline
 2192  &  2198 &  2209	&  2254  &  2256  &  2259  &  2264  &  2372  &	2381  &  2391	\\
 2418  &  2452 &  2457	&  2461  &  2480  &  2485  &  2506  &  2516  &	2526  &  2528	\\
 2532  &  2552 &  2556	&  2565  &  2576  &  2580  &  2586  &  2604  &	2611  &  2629	\\
 2637  &  2668 &  2680	&  2681  &  2684  &  2702  &  2713  &  2720  &	2722  &  2724	\\
 2726  &  2733 &  2739	&  2747  &  2748  &  2752  &  2755  &  2761  &	2763  &  2764	\\
 2768  &  2769 &  2777	&  2787  &  2793  &  2794  &  2796  &  2801  &	2805  &  2807	\\
 2812  &  2814 &  2815	&  2816  &  2818  &  2820  &  2821  &  2822  &	2823  &  2826	\\
 2834  &  2840 &  2842	&  2844  &  2846  &  2847  &  2849  &  2850  &	2852  &  2854	\\
 2857  &  2858 &  2864	&  2866  &  2869  &  2874  &  2875  &  2877  &	2878  &  2886	\\
 2890  &  2891 &  2892	&  2899  &  2900  &  2903  &  2906  &  2907  &	2910  &  2911	\\
 2923  &  2935 &  2979	&  2980  &  2994  &  3000  &  3001  &  3012  &	3029  &  3036	\\
 3041  &  3043 &  3050	&  3063  &  3066  &  3069  &  3082  &  3084  &	3093  &  3112	\\
 3116  &  3125 &  3134	&  3136  &  3178  &  3189  &  3264  &	     &	      & 	\\
\hline
\hline
\end{tabular}
\end{flushleft}
\end{table}

Table 6 yields those objects which are members only in one
or two streams; the term ``yes'' denotes that an object is member of a
given stream, the term ``no'' denotes that an object is not member of a
given stream.

\begin{table}
\caption{Objects of meteor streams maximally in two of the three cases.}
\begin{flushleft}
\begin{tabular}{cccc}
\hline
\hline
object & case 1 & case 2 & case 3 \\
\hline
2248   &   no	&  no	 &   yes \\
2298   &   yes	&  no	 &   yes \\
2371   &   no	&  yes	 &   yes \\
2523   &   no	&  yes	 &   yes \\
2547   &   yes	&  no	 &   no  \\
2599   &   yes	&  no	 &   no  \\
2727   &   no	&  yes	 &   no  \\
\hline
\hline
\end{tabular}
\end{flushleft}
\end{table}

While in the cases 1 and 2 the number of objects belonging only to one
of the streams to the number of objects belonging to the both streams is 3/127
$\approx 2.4 \%$, all three cases yield similar ratio 7/127
$\approx 5.5 \%$. It is evident that this number may even increase when
other areas $\Omega$ are used.

All three cases yield $\lambda_{\odot} \in ( 165^{\circ}, 265^{\circ} )$.
Distribution in $\lambda_{\odot}$ is not approximable by normal distribution.
The relation between quantities of different distributions
in a form of linear regression is not wise. This holds, as a special case,
between an orbital element and $\lambda_{\odot}$ -- even
in a small interval $\Delta \lambda_{\odot}$ (as for larger interval, the
procedure may even not be used due to the fact that periodicity is not secured).
Evident approximate relation exists between $\pi$ and $\lambda_{\odot}$ due to the
well-known relation among $\pi - \Omega - \lambda_{\odot}$; however, we
again emphasize that density functions for $\pi$ and $\lambda_{\odot}$
are different. However, in order to be able to compare our Taurid stream
with that obtained in the past, we present the following linear regressions:
\begin{eqnarray}\label{2}
q [AU] &=& (0.096 \pm 0.018) ~-~ (0.791 \pm 0.811) 10^{-4} \lambda_{\odot} [^{\circ}] ~ , \\ \nonumber
\pi [^{\circ}]	&=& (- 12.065 \pm 4.710) ~+~ (0.7563 \pm 0.0213) \lambda_{\odot} [^{\circ}] ~ ,
\end{eqnarray}
for the case 1,
\begin{eqnarray}\label{3}
q [AU] &=& (0.094 \pm 0.018) ~-~ (0.671 \pm 0.816) 10^{-4} \lambda_{\odot} [^{\circ}] ~ , \\ \nonumber
\pi [^{\circ}]	&=& (- 18.410 \pm 5.246) ~+~ (0.7826 \pm 0.0237) \lambda_{\odot} [^{\circ}] ~ ,
\end{eqnarray}
for the case 2, and,
\begin{eqnarray}\label{4}
q [AU] &=& (0.087 \pm 0.017) ~-~ (0.363 \pm 0.791) 10^{-4} \lambda_{\odot} [^{\circ}] ~ , \\  \nonumber
\pi [^{\circ}]	&=& (- 16.756 \pm 4.887) ~+~ (0.7759 \pm 0.0221) \lambda_{\odot} [^{\circ}] ~ ,
\end{eqnarray}
for the case 3. Comparison of Eqs. (2) -- (4) with the results of
Porub\v{c}an and \v{S}tohl (1987 -- Table 1) shows that our Taurid meteor complex
is different from that obtained on the basis of D-criterion of Southworth
and Hawkins (1963).

We must stress again that criterion suggested in paper I (in section 5),
and applied in this paper, is of purely mathematical character,
based on mathematical statistics. Our criterion is not of physical character,
as was interpreted in Neslu\v{s}an (1996).

\section{Conclusion}
We have applied methods developed in papers I and II to meteor streams of
comet Encke.
We have found, in a simple physical way, theoretical radiants
which should correspond to comet Encke as a parent body. The coincidence
with the reality is not very bad.

Taurid meteor complex, based on the data of photographic orbits, was found
for various areas in phase space. Comparison of the results based on
statistical method of multidimensional density function is not consistent
with that
obtained on the basis of Southworth and Hawkins D-criterion.
The difference will even increase with the increasing number of photographic
meteors.

We can conclude that our definition of the Taurid meteor complex
corresponds to the multidimensional normal distribution in phase space
of orbital elements. As an example,
we have used three different areas $\Omega$.
Physical problem is to make a decision which kind of area $\Omega$
is the most reasonable.

\acknowledgements
Special thanks to the firm ``Pr\'{\i}strojov\'{a} technika, spol. s r. o.''.
This work was partially supported by Grants VEGA No. 1/4304/97 and
1/4303/97.

\end{document}